# Behavioral an real-time verification of a pipeline in the COSMA environment

Jerzy Mieścicki, Wiktor B. Daszczuk[*]

*Institute of Computer Science, Warsaw University of Technology,
Nowowiejska 15/19, PL 00-665 Warsaw, Poland*

**Abstract**

The case study analyzed in the paper illustrates the example of model checking in the COSMA environment. The system itself is a three-stage pipeline consisting of mutually concurrent modules which also compete for a shared resource. System components are specified in terms of Concurrent State Machines (CSM) The paper shows verification of behavioral properties, model reduction technique, analysis of counter-example and checking of real time properties.

## 1. Introduction

In [1] we have described the functional model of a system for processing of consecutive portions of data (or messages) submitted to its input. Each message goes through the three stages of processing which is reflected in the system structure (Fig. 1). The system is a three-stage pipeline consisting of three modules that operate concurrently and asynchronously, in a sense that there is no general, common synchronizing process or mechanism. Moreover, two out of three modules compete for the access to the common resource, which is accessed also by some other (unspecified) agents from the system environment. This calls for the verification if the cooperation among system components is *correct*. Indeed, due to potential coordination errors the system can get deadlocked, messages can be lost or duplicated etc. After the behavior is proved correct, some real-time performance features may be formally analyzed: minimal and maximal time of given actions, time intervals between events etc.

It is known that in the case of asynchronous and concurrent systems behavioral errors are extremely hard to discover, identify and correct using typical debugging and testing procedures. Therefore, we have applied a formal procedure of *model checking* [2-5], using the software tool called COSMA [6],

[*]Corresponding author: *e-mail address*: W.Daszczuk@ii.pw.edu.pl

implemented in the Institute of Computer Science, Warsaw University of Technology.

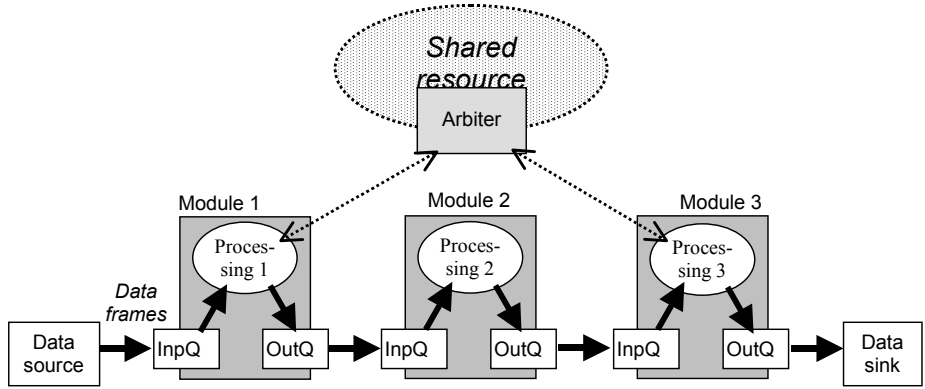

Fig. 1. Flow of data in a three-module pipeline with a shared resource

Model checking is based on the following principle. Given the finite state model *M* of system behavior and property (requirement) *p* to be checked, one has to check if *p holds for M*. Usually, there is a set of techniques and algorithms (making together the model-checking environment or tool) designed for this purpose. This is a designer's job to formulate properties to be evaluated: usually the verification involves a set of model checking experiments with several properties $p_i$. Additionally, if the given property does not hold for *M*, then so-called counterexample is provided which allows to identify the sequence of states (or events) that results in this negative evaluation. This helps to identify and correct the cause of an error.

The main limitation the model checking faces is the *exponential explosion* of model's state space size along with the increase of the number of finite state system components and their individual state spaces. So, an extensive research is being done on various techniques that can help to manage the problem. First, multiple forms of *reduction* of state space are proposed, aimed at removal of the states and transitions which are irrelevant w.r.t. the evaluation of a given formula. The other approach is to calculate the state space just during the evaluation, as one can expect that in order to obtain the outcome of the evaluation only the *bounded model* will do. Still another technique consists in *compositional model checking*, where some individual parts of a system (of more acceptable size) are subject to an exhaustive state space search while the conclusion as to the behavior of a whole system is reached by some logical reasoning. Unfortunately, most ideas of reduction found in literature (e.g. [7,8]) usually cannot be applied for Concurrent State Machines. This model admits *coincident* execution of actions rather than their interleaving, while most finite state models assume just the interleaved executions. Also, other known forms of reduction (e.g. *slicing* and *abstraction* [9-11]) make the use of specific properties of programs and can not be applied directly to more abstract CSM models.

In this paper we briefly describe three techniques used in a COSMA-style methodology of system verification. First, we will analyze the system behavior step-by-step, using so-called multi-phase reduction [12,13] which exploits some compositional features of the CSM model [14]. As a result, the system which (as naively estimated) may have as much as $4*10^{14}$ states is finally reduced to a model of 323 states and 1406 edges, easily representable and algorithmically checkable in a split second. Then, as some properties proved to be evaluated to *false*, we illustrate how the counterexample can be obtained and analyzed. Finally, using timed version of the model, we present how real time dependencies may be analyzed.

## 2. Two-phase procedure of obtaining the reduced reachability graph

Let us recall the basic facts about the CSM model of a pipeline, described in more details in [1]. It consists of three complex modules and three individual components (data source, data sink and the arbiter) common to the whole system. Each module can be internally subdivided into six components (see also left-hand part of Fig. 2). In total, this makes a set of 21 cooperating components. For each of them, a separate (finite state) CSM model has been developed, aimed at specifying its behavior as well as the communication to/from its communication partners. The goal was to obtain the large system's behavioral model or a graph of reachable system states, containing all the reachable states and possible execution paths among them. Then, some temporal formulas representing desirable behavioral properties of the system have been evaluated (*true* or *false*).

In [1] the emphasis was put on the specification of components and temporal properties, while the technique of obtaining the product of all the components was not analyzed. Now we proceed to the method of determining the system's behavioral model that can (to an extent) help to cope with problems of the graph size. The main idea devoted to is the following. In order to obtain a system behavioral model, one has to perform the product ($\otimes$) of CSM models of system components. This operation is associative and commutative. Associativity supports the important compositional property. Now, if we have – for instance – a system $Z = \{m, n, p\}$ of three components, then (due to the associativity) we can obtain the behavioral model either immediately, as a 'flat' product $\otimes Z = m \otimes n \otimes p$ or in two steps: first computing the local product of some subsystem e.g. $r = m \otimes n$, then $\otimes Z = r \otimes p$. Meanwhile, before the second, final product is obtained, we can apply some reduction procedure to the partial product *r*. While the associativity of the product applies to other finite state models as well, this reduction makes the use of intrinsic features of CSM model itself. If machines *m* and *n* do communicate intensively with each other – it may result in a considerable reduction of a total computational effort, necessary for the computation of ⊗Z.

Below, we show how this general rule applies to our system of 21 components, briefly recalled above. We will proceed in the two steps or *phases*.

Generally, each phase consists in the selection of some subsystem, obtaining its CSM product and removing the *irrelevant* states and edges from it. However, one has to decide first which elements of the model are *relevant* ones and therefore have to be preserved. Relevant – in this sense – are the selected *output symbols* (produced by individual system components) and thus also the system states in which these symbols are generated. Typically, among relevant symbols are:
1. symbols that are referred to in the temporal formulas to be evaluated,
2. symbols that should be preserved for designer's convenience, e.g. because they make the complex behavior more readable,
3. symbols that are necessary from the viewpoint of the communication among the currently reduced subsystem and remaining components[1].

The former two groups of symbols are decided upon by the designer while the latter one is determined by the specification of system components. Assume that in our case the relevant symbols of types 1 and 2 above are the following ones:

$$msg\_1, msg\_4, doProc\_1, doProc\_2, doProc\_3$$

In order to obtain the 'phase-1' model of our example system we perform the following procedure:

**Phase-1**
1. take a subsystem, consisting of the six components of module #1 (Fig. 2),
2. compute its CSM product,
3. reduce it, leaving as the relevant output symbols the following ones:
   – all the output symbols (from the subsystem) which are 'watched for' by the subsystem communication partners (i.e. the *Arbiter*, *Trsm_0*, *Rcv_2*),
   – symbols from the set selected above, which are produced within module #1 (this case: *doProc_1*),

Let the product reduced this way be called *Subsystem_1*,

4. repeat the above for modules #2 and #3 obtaining *Subsystem_2* and *Subsystem_3* (respectively),
5. Substitute subsystems 1, 2, 3 in the place of just processed components.

---

[1]Note that among the 'remaining components' can be also the additional, auxiliary automata (e.g. *Invariant*, in our case) necessary for expressing the properties under checking.

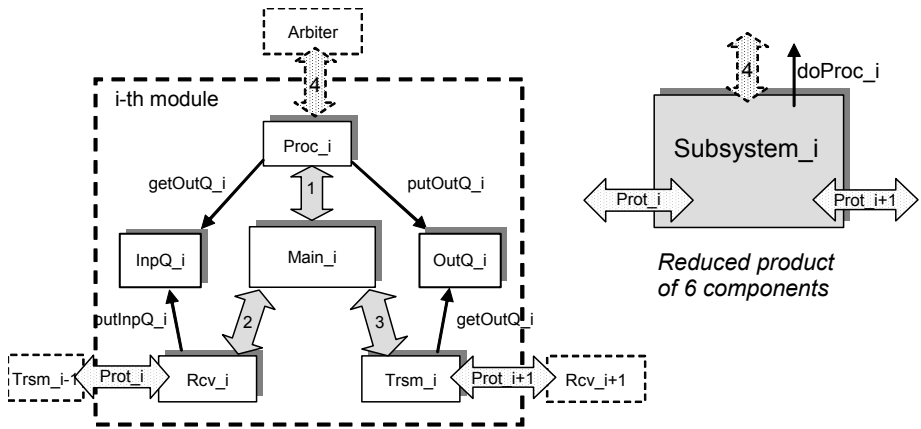

Fig. 2. 'Local' product of a single module

This way we obtain the phase-1 structural model, in which subsystems 1, 2 and 3 are replaced by single automata. It is noteworthy that for *Subsystem_1*:
– Cartesian product of its six components has 24300 states,
– CSM product (before reduction) has 24 reachable states and 31 edges,
– after reduction, *Subsystem_1* is a graph of 10 states and 16 edges.

For *Subsystem_3*, the situation is analogous. As an illustration, the reduced CSM product of six components making *Subsystem_3* (*Main_3, Rcv_3, Trsm_3, Proc_3, InpQ_3, OutQ_3*) is shown in Fig. 3. At no surprise, it has 10 states and 16 edges, the same as *Subsystem_1*. For *Subsystem_2* (not shown), there are as few as 7 states and 12 edges.

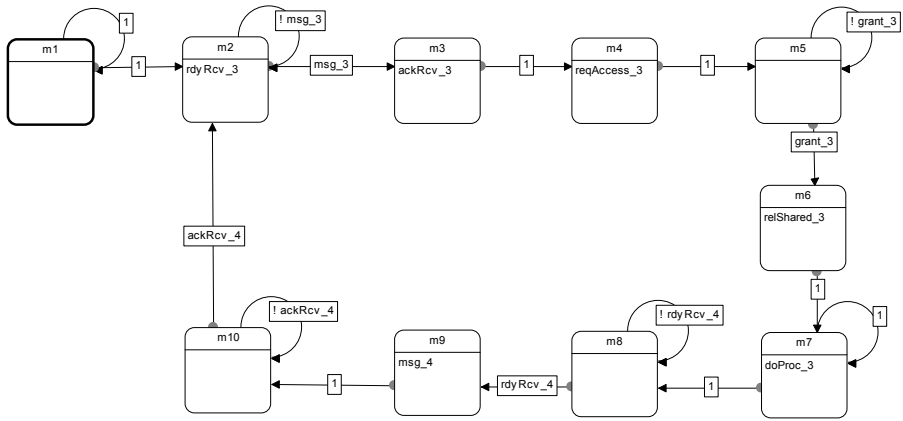

Fig. 3. CSM model of *Subsystem_3* (*reduced* product of six components of module #3)

Now, the analogous procedure can be applied again to the structural elements of the reduced model, for instance:

**Phase-2**
- Apply the procedure to a subsystem consisting of *Subsystem_1* and *Trsm_0*, preserving all the output symbols which are 'watched for' by the communication partners (i.e. *Arbiter* and *Subsystem_2*) and symbols needed for temporal formulas to be evaluated (this case: *doProc_1* and *msg_1*);
- and to a subsystem consisting of *Subsystem_3* and *Rcv_4*, preserving 'watched for' symbols (i.e. *Arbiter* and *Subsystem_2*) and symbols for model checking (this case: *doProc_3* and *msg_4*);
- finally substitute *Syst_1_Trsm_0* and *Syst_3_Rcv_4* in the place of just processed components.

This way we obtain the phase-2 structural model as in Fig. 4. Notice that the phase-2 system now consists of *four* components (instead of 21 components of phase-0 structural model), each of significantly reduced size. This 'downsizing' the model can be continued, but each time the reduction is performed certain conditions have to be met [12] so that the reduction is not necessarily guaranteed. Nevertheless, in practice the degree of reduction can be substantial.

Let the CSM product of the system from Fig. 4 be called *New_System* and serve as the new behavioral model in which the temporal requirements are evaluated. *New_System*, obtained again with the COSMA Product Engine, has 323 states and 1406 edges and is expected to preserve at least these functional properties of the original, flat version which can be expressed in terms of symbols *msg_1, msg_4, doProc_1, doProc_2, doProc_3*.

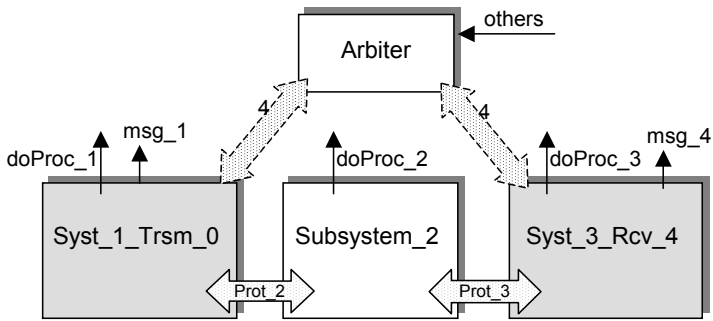

Fig. 4. Phase-2 structural model of the system

### 3. Verification of the reduced model

To sum up, now we have two behavioral models of the same example system:
- *Flat-product* (CSM product of 21 components, obtained as described in [1]) which had 8284 states and 34711 edges,
- *New_System*, obtained in the above two-phase reduction procedure, with 323 states, 1406 edges.

Both models have been verified in the COSMA environment. As in [1], an additional automaton *Invariant* was determined to conveniently specify the verified properties. The checked properties were the following:
- <u>Safety 1</u>, saying – informally – that the number of messages within the pipeline never exceeds its capacity and the number of messages leaving the pipeline never exceeds the number of messages entering it,
- <u>Liveness 1</u>, saying – informally – that for any system state it is possible that the pipeline *eventually* would get empty,
- <u>Liveness 2</u>, saying – informally – that for any system state it is possible that the pipeline *eventually* would get full.

Experiments have been performed on PC computer with 800MHz processor and 512 MB RAM. The results are summarized in Table 1.

Table 1. Summary of experiments

|  | Flat model | | Reduced model | |
| --- | --- | --- | --- | --- |
|  | Result | Evaluation time | Result | Evaluation time |
| Safety 1 | true | 17 s | True | < 1 ms |
| Liveness 1 | false | 54 s | False | < 1 ms |
| Liveness 2 | false | 4 min 40 s | False | 60 ms |

Notice that both formulas referring to the liveness have been evaluated *false* in both (flat and reduced) models. This negative result means that the system may enter such a state (states) that – from this state on – the pipeline is never empty again (i.e. it never terminates the processing of messages) or is never able to process three messages at once, which it was designed for.

The differences in evaluation times are really noteworthy: in all cases the ratio of $10^3$-$10^4$ in favor of reduced model was achieved, even though in terms of state space size the reduced model is only approximately 25 times less than the flat one.

Finally, the model verification can be summarized as follows;
- The system itself performs wrong: there must be a synchronization bug in the specification of components. This calls for the analysis of a counterexample.
- The reduced model well preserves the relevant properties of the primary, flat one. Indeed, each case the same temporal formula was evaluated the same way (*true* or *false*) in both models.
- The multi-phase reduction method provides a significant gain in the evaluation time, even greater than the savings in the state space itself.
- The advantages of the evaluation algorithm used in the COSMA tool have been also confirmed. The algorithm terminates the evaluation as soon as the result (*true*, *false*) is certainly determined. It is why the evaluation times of rather similar formulas (1) and (2) differ by a few dozen of times.

## 4. Analysis of a counterexample

In the case of negative evaluation, the TempoRG checker [15-17] produces a counterexample. Often, it is a path (a sequence of states) in the reachability graph that leads to the state where it was decided that the temporal formula is to be certainly evaluated false. In the case of more complex temporal formulas involving several operators, the counterexample can be a tree [15], showing which particular part of the formula (a sub-formula) is responsible for the negative result. Tracing the consecutive states along the counterexample, the designer is able to identify the synchronization bug.

However, in the case of *reduced* models, the model states can be unreadable. As a result of reduction, some states are eliminated, the remaining ones are usually renamed etc., so that the analysis of counterexample should be based on the sequences of symbols (events) produced by the system instead of on sequences of states.

The evaluation of both formulas representing the liveness condition yields the same counterexample, presented in Fig. 5. The counterexample itself pretends to be a CSM, in order to enable the use of animation feature of COSMA tool. Using it, one can trace the states of individual components (and their change) corresponding to consecutive states of an counterexample. Also, some additional symbols (not used in 'regular' CSM) are introduced as first elements of states' output field. @ marks the starting state of the formula (in this case it is the system initial state) while F and G stand for the operators of sub-formulas (*G* stands for **AG** and *F* stands for **AF**).

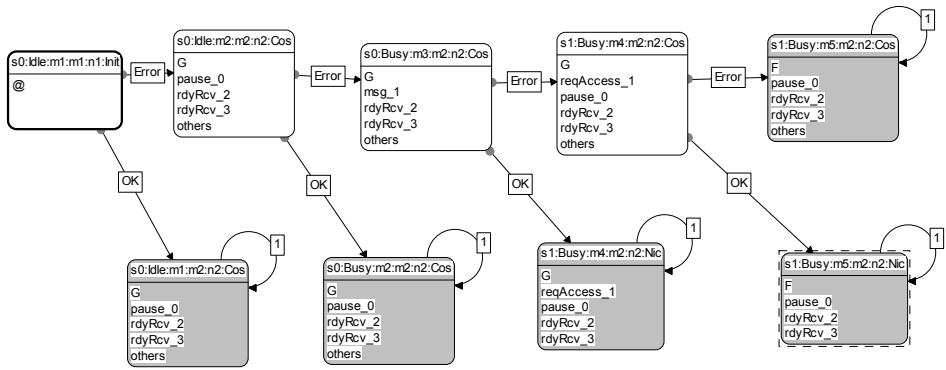

Fig. 5. Counterexample to the formula **AG AF in** *Invariant.s3*

The counterexample is constructed as follows:
- it begins in the starting state of evaluation (the initial state in this example),
- it contains sub-paths responsible for sub-formulas (which may produce a tree-like counterexample),
- for four states two successors are shown: one which leads towards an erroneous state (in which the error is possible, transition labelled with *Error*), the other one which leads to a 'proper' state (transition labelled with *OK*),
- the fifth state in an upper sequence, namely *s1:Busy:m5:m2:n2:Cos* is referred to as a *Trap*. The rules of constructing counterexamples [24] say that this state is a representative of so-called Ending Strongly Connected Subgraph (ESCS) of states in which the most nested formula (**in** *Invariant.state*; *state* ∈{*s0,s3*}) is not satisfied. When the system falls into one of these states, the error is inevitable (the desired state *state* of *Iinvariant* is never reached).

The analysis starts with finding the last one of states (in the sequence) that has two outgoing transitions: one labelled *OK* and the other labelled *Error*. This state is referred to as a *Checkpoint*. In the example, it is (*s1:Busy:m4:m2:n2:Cos*), with two successors: (*s1:Busy:m5:m2:n2:Nic*) as a 'proper' state and (*s1:Busy:m5:m2:n2:Nic*) as a 'wrong' one. This time, the 'wrong' state is actually the *Trap* itself, but often it is only the initial state of a sequence of states which *inevitably* ends in a trap. Analysis of signals generated in the triangle *{Checkpoint, its 'proper' successor, its 'wrong' successor}* reveals the nature of error. We see that in the *Checkpoint* a request of access to the shared resource is generated (signal *req_Access_1*), and the resource is granted to another user (signal *others* is present). For this state, its 'proper' successor does not produce *others*, while in the *Trap* the symbol *others* is still present. So, *OK*-labelled transition (to a 'proper' state) is executed only if the signal *others* is withdrawn, otherwise the system chooses a transition to a 'wrong' state which leads to the *Trap*. In other words, the error is inevitable, if the request (*req_Access_1*) is issued while other users do use the shared resource.

Actually, in the system the two-state dead-end subgraph (causing a livelock of the whole system) can be found. The system performs incorrectly because *reqAccess_1* is not stored. Recall that in the CSM framework no implicit buffering of events is assumed: this should be provided by the model itself, e.g. by an additional (e.g. two-state) buffering component or by a simple modification of *Proc_1*. The same conclusion refers to the third module which accesses the shared resource as well. Both modules (#1 and #3) have been easily corrected and positively verified.

Finally, we may add that the flat product of the *corrected* system has 8086 states, 33588 edges instead of 8284 states and 34711 edges of the (incorrect) flat product discussed in [1]. This confirms the observation that the better the synchronization is, the less is the behavioral model of a system.

## 5. Real-time dependencies

Now we may convert automata to TSCM (Timed CSM, derived from CSM as Timed Automata [18,19]) by adding time constraints and clock resets on some transitions in automata *Proc_i* and control units *Main_i* (Fig. 2). All time dependencies are shown as multiples of a basic time period, a *tick*. The constraints in *Proc_i* (Fig. 6) inform what is the minimal time of processing (*tim1*: by the constraint on the transition outgoing from the state *useshared*) and the maximal time (*tim2*: the constraint on self-loop of the state *useshared*). The constraints are based on a clock $T_i$ local to *Proc_i*. The fixed time of staying in states in *Main_i* models delays in control unit. The clock is reset every time the automaton enters *useshared*. The constraints guarantee that the time of using a shared resource is finite. The constants *tim1_i* and *tim2_i*, *tim1_i* < *tim2_i*, may be specific to subsystems 1,2 and 3. Auxiliary automaton which guarantees finite time of using the resource by *others* must be modelled (instead of the external signal *others*). Also, maximal time of a time period between generation of items should be specified.

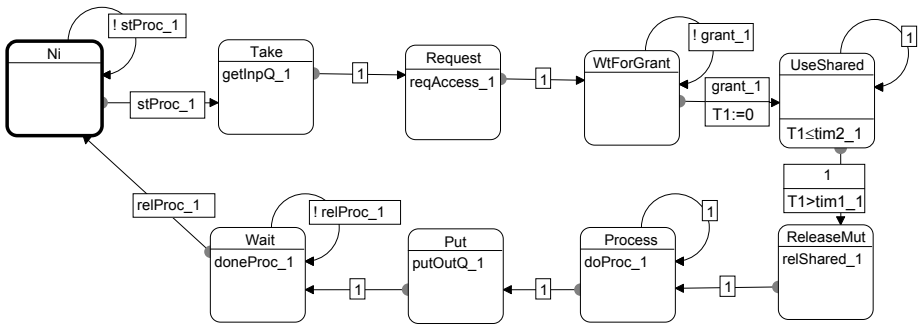

Fig. 6. Timed automaton *Proc_1*

Unfortunately, TCSM does not specify the succession relation unambiguously. The RCSM (Region CSM automaton) may be calculated from the product TCSM, following the rules given in [20]. Storing a timed automaton in the RCSM form allows the verification system to compute its products with various testing automata. For this purpose, rules for multiplication of RCSM automata were developed [20].

Based on the RCSM state space, a testing automaton may be constructed, as shown in Fig. 7. This automaton checks if a time period between two items on output of the whole system is *<0,1)*, *<1,2)*, *<2,3)* ... ticks. If we impose minimal and maximal time on the system, states violating the limits should produce the *error* signal (period *<1* or *>4* in this case).

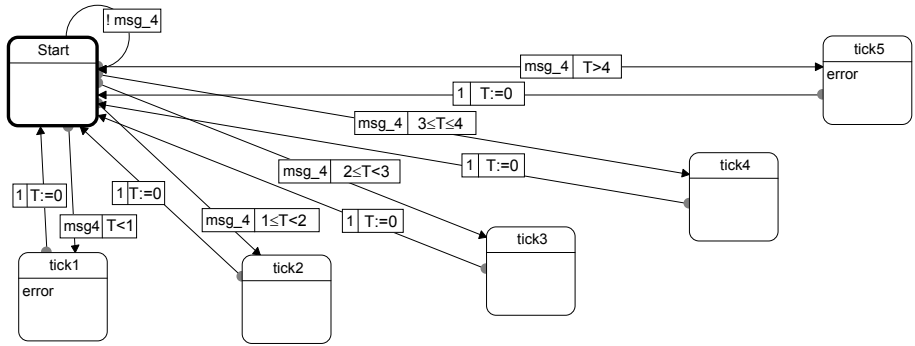

Fig.7. Testing timed automaton

The presented verification should be completed by former tests for liveness and safety (but in the RCSM state space), because time constraints may change the behavior of the system and the results obtained for CSM may be no longer valid.

## 6. Conclusions

The advantage of (Timed) Concurrent State Machines formalism is that in order to understand (or even to design) the behavioral specification of a system component one has to be familiar with only a few elementary notions: a state, a transition, an atomic symbol, a Boolean formula, a time constraint. Generally, the semantics of an individual CSM is not far from the conventional finite state machines or basic UML's state diagram. However, given a collection of such CSM components, one can select a subsystem and obtain its product, representing (in one, large graph) all possible subsystem's executions or runs. Consequently, the model of a system can be subject to formal model checking methods and techniques. This advantage is not provided by standard specification methods based on UML.

Moreover, as we have shown, the COSMA software environment supports the additional functional features, like stepwise model reduction, defining behavioral invariants, imposing time dependencies etc., as well as the means for the analysis of counterexamples. This makes the (Timed) Concurrent State Machines (and COSMA tool) a good candidate for a convenient framework for preliminary specification of concurrent, reactive systems. Once verified and corrected, such a specification can be refined, enhanced and otherwise developed in other professional software development environments. Moreover, if some components are to be *hardware* – implemented (which is often the case

*Behavioral an real-time verification of a pipeline ...* 265in embedded systems), the automata-like CSM specification is also close to common forms of behavioral specification of sequential circuits.

This work has been supported by grant No.7 T 11 C 013 20 from the Polish State Committee for Scientific Research (Komitet Badań Naukowych).

# References

[1] Mieścinki J., Czejdo B., Daszczuk W.B., *Model checking in the COSMA environment as a support for the design of pipelined processing*. Proc. European Congress on Computational Methods in Applied Sciences and Engineering ECCOMAS 2004, Jyväskylä, Finland, (2004).
[2] Clarke E.M., Grumberg O., Peled D.A., *Model Checking*. MIT Press, (2000).
[3] Berard B. (ed.) et al., *Systems and Software Verification: Model-Checking Techniques and Tools*. Springer Verlag, (2001).
[4] Peled D.A., *Software Reliability Methods*. Springer Verlag, (2001).
[5] McMillan K.L., *Symbolic Model Checkin*. Kluwer Academic Publishers, (1993).
[6] COSMA: www.ii.pw.edu.pl/cosma/
[7] Gerth R., Kuiper R., Peled D.A., Penczek W., *A Partial Order Approach to Branching Time Logic Model Checking*. Information and Computation. 150(2) (1999) 132.
[8] Wolper P., Godefroid P., *Partial-Order Methods for Temporal Verification, in Proc. of CONCUR '93*. Lecture Notes in Computer Science, Springer-Verlag, New York, 715 (1993) 233.
[9] Corbett J.C., Dwyer M.B., Hatcliff J., Laubach S., Păsăreanu C.S., Robby Zheng H., *Bandera: Extracting Finite-state Models from Java Source Code*. in Proc. of the 22nd International Conference on Software Engineering, ICSE 2000, June 4-11, Limerick, Ireland. ACM, (2000) 439.
[10] Păsăreanu, C., S., Dwyer, M., B., Visser, W., Finding Feasible Counter-examples when Model Checking Abstracted Java Programs, in Proc. 7th International Conf. on Tools and Algorithms for the Construction and Analysis of Systems, TACAS 2001, Genova, Italy, April 2-6, 2001, Lecture Notes in Computer Science vol. 2031, p. 284.
[11] Harman M., Danicic S., *A New Approach to Program Slicing*. in Proc. of 7th International Software Quality Week, San Francisco, (1994).
[12] Mieścicki J., *Multi-phase model checking in the COSMA environment. Institute of Computer Science*. WUT, Research Report, 14 (2003).
[13] Mieścicki J., Czejdo B., Daszczuk W.B., *Multi-phase model checking in the COSMA environment as a support for the design of pipelined processing*. Institute of Computer Science, WUT, Research Report, 16 (2003).
[14] Mieścicki J., *Concurrent State Machines, the formal framework for model-checkable systems*. Institute of Computer Science, WUT, Research Report, 5 (2003).
[15] Daszczuk, W. B., *Critical trees: counterexamples in model checking of CSM systems using CBS algorithm*. Institute of Computer Science, WUT, Research Report, Warsaw, 8 (2002).
[16] Daszczuk W.B., *Verification of temporal properties in concurrent systems*. Ph. D. Thesis, Faculty of Electronics and Information Technology, Warsaw University of Technology, (2003).
[17] Daszczuk W.B., *Temporal model checking in the COSMA environment (the operation of TempoRG program)*. Institute of Computer Science, WUT, Research Report, 7 (2003).
[18] Alur R. Dill D.L., *A Theory of Timed Automata*. in Theoretical Computer Science, 126 (1994) 183.
[19] Alur R., *Timed Automata*. in 11th International Conference on Computer-Aided Verification, LNCS 1633, Springer-Verlag, (1999) 8.
[20] Daszczuk W.B.,. *Timed Concurrent State Machines*. Institute of Computer Science, WUT, Research Report, 27 (2003).